\documentclass[aps,pra,reprint,superscriptaddress]{revtex4-1}
\usepackage[utf8]{inputenc}

\usepackage{graphicx}  
\usepackage{xcolor}
\usepackage{bm}        

\usepackage{amssymb} 
\usepackage{amsmath}
\usepackage{amsthm}
\usepackage{amsfonts}
\usepackage{mathtools}
\usepackage{ragged2e}
\usepackage{tikz}
\usepackage{hyperref}

\usepackage[capitalize]{cleveref}
\crefname{section}{Sec.}{Secs.}

\usepackage{bbm}
\usepackage{todonotes}

\theoremstyle{plain}

\theoremstyle{definition}

\newcommand{\ket}[1]{\left|#1\right\rangle}
\newcommand{\bra}[1]{\left\langle#1\right|}

\begin{document}
	\title{Simulating X-ray absorption spectroscopy of battery materials on a quantum computer}
	
	\author{Stepan Fomichev}
	\thanks{These authors contributed equally.\\
		stepan.fomichev@xanadu.ai\\
		kasra.hejazi@xanadu.ai}
	\affiliation{Xanadu, Toronto, ON, M5G2C8, Canada}
	\author{Kasra Hejazi}
	\thanks{These authors contributed equally.\\
		stepan.fomichev@xanadu.ai\\
		kasra.hejazi@xanadu.ai}
	\affiliation{Xanadu, Toronto, ON, M5G2C8, Canada}
	\author{Ignacio Loaiza}
	\affiliation{Xanadu, Toronto, ON, M5G2C8, Canada}
	\author{Modjtaba Shokrian Zini}
	\affiliation{Xanadu, Toronto, ON, M5G2C8, Canada}
	\author{Alain Delgado}
	\affiliation{Xanadu, Toronto, ON, M5G2C8, Canada}
	\author{Arne-Christian Voigt}
	\affiliation{Volkswagen AG, Berliner Ring 2, 38440 Wolfsburg, Germany}
	\author{Jonathan E. Mueller}
	\affiliation{Volkswagen AG, Berliner Ring 2, 38440 Wolfsburg, Germany}
	\author{Juan Miguel Arrazola}
	\affiliation{Xanadu, Toronto, ON, M5G2C8, Canada}

	\begin{abstract}
		X-ray absorption spectroscopy is a crucial experimental technique for elucidating the mechanisms of structural degradation in battery materials. However, extracting information from the measured spectrum is challenging without high-quality simulations. In this work, we propose simulating near-edge X-ray absorption spectra as a promising application for quantum computing. It is attractive due to the ultralocal nature of X-ray absorption that significantly reduces the sizes of problems to be simulated, and because of the classical hardness of simulating spectra. We describe three quantum algorithms to compute the X-ray absorption spectrum and provide their asymptotic cost. One of these is a Monte-Carlo based time-domain algorithm, which is cost-friendly to early fault-tolerant quantum computers. We then apply the framework to an industrially relevant example, a CAS(22e,18o) active space for an O-Mn cluster in a Li-excess battery cathode, showing that practically useful simulations could be obtained with much fewer qubits and gates than ground-state energy estimation of the same material.
	\end{abstract}

	\maketitle

	\section{Introduction}
	
	The commercial development of durable cathodes with high energy density is one of the key challenges facing the battery manufacturing industry. Li-excess cathodes are a promising candidate for achieving high capacity by virtue of over-saturating the material with a surplus of lithium ions \cite{lu2002understanding,lu2002synthesis,saubanere2016intriguing,tran2008mechanisms,zhang2022pushing,hong2015lithium,hy2016performance,radin2017narrowing}. However, the performance of such materials rapidly degrades with subsequent charge-discharge cycles \cite{johnson2008synthesis,bettge2013voltage}, likely due to irreversible structure degradation triggered when a certain charging level is reached. 
	
	There are a number of competing hypotheses as to the cause of the degradation, such as anionic (oxygen) redox leading to dimerization of oxygen atoms \cite{seo2016structural,luo2016charge,koga2013different,koga2014operando,house2023delocalized}, or oxidation of manganese ions from the $4^+$ to the $7^+$ oxidation state \cite{kalyani1999lithium,ohzuku2011high,radin2019manganese}. If the formal oxidation states of the constituent atoms could be accurately determined, one of these hypotheses could be confirmed and strategies to avoid degradation could be devised. 
	
	The key source of evidence regarding oxidation states in Li-excess cathodes is X-ray absorption spectroscopy (XAS) \cite{radin2019manganese,house2023delocalized}. XAS is a well-established and widely used experimental technique that provides unique access to the local geometric and electronic structure of molecules and materials, including the valence and oxidation states of individual atomic species \cite{de2001high}. However, this method relies on matching the observed spectrum with that of reference compounds with known oxidation states: whenever these are unavailable, ab initio simulations are required to extract information.
	
	Much like for ground-state energy calculations, a great variety of ab initio methods have been devised to simulate X-ray absorption spectra. However, in practice it has turned out to be much harder to identify efficient heuristics for computing spectra than ground states. A similar observation can be made on theoretical grounds: while there are results suggesting that ground state calculations can in principle be performed efficiently, for example with tensor networks \cite{hastings2007entropy}, no similar results have emerged for excited states relevant for simulating spectra. Combined with the ultralocal nature of near-edge X-ray spectra that requires only the simulation of small clusters (nearest-neighbour shell), these computational challenges associated with XAS simulation suggest that this problem could be a natural fit for a quantum computer. 
	
	\begin{figure*}[t]
		\centering
		\includegraphics[width=0.9\linewidth]{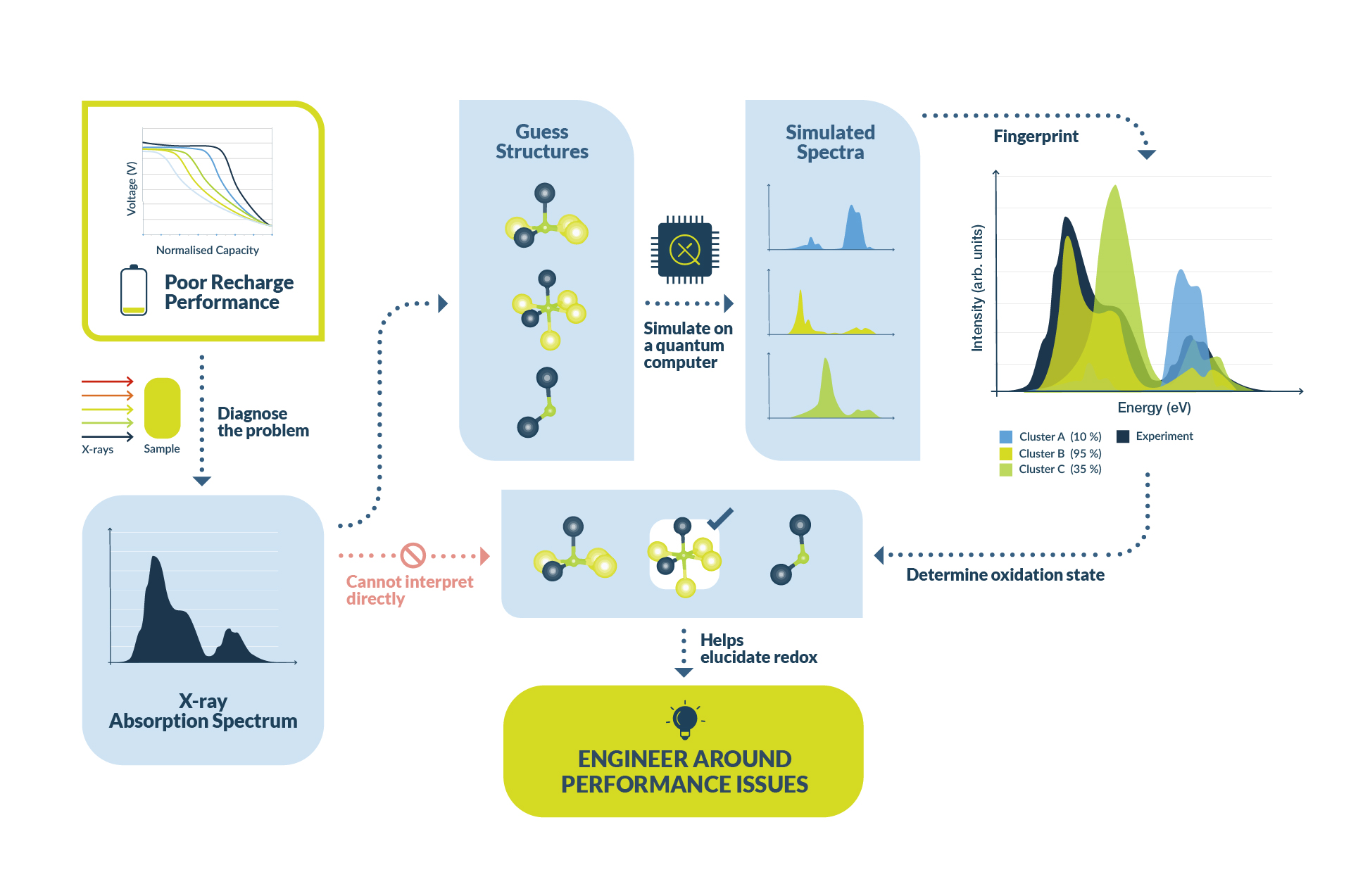}
		\caption{In order to overcome poor recharge performance of Li-excess batteries, it is necessary to understand the redox processes occurring in the cathode. XAS is the key experimental approach to gaining this information, but simulations are required to interpret the results. Quantum computing can help with this challenge by providing accurate simulations of X-ray spectra that are inaccessible to classical simulation methods. Through spectral fingerprinting, these simulations yield information such as oxidation states, that can point the way towards better cathode materials and battery designs.}
		\label{fig:hero}
	\end{figure*}
	
	In this work, we identify the simulation of X-ray absorption spectra as a promising application of quantum computing. We summarize the proposed application in \cref{fig:hero}. For a concrete example, we focus on quantum algorithms for obtaining the spectra of clusters that arise during delithiation of Li-excess cathodes. We start by briefly describing the X-ray absorption process and the key ingredients for modelling it in Li-excess materials, highlighting the limitations of existing classical methods and the aspects that make XAS a good fit for quantum computing. We then propose three techniques -- the frequency-domain Green's function method, the quantum phase estimation (QPE) sampling method, and the newly-developed Monte-Carlo-based time-domain method  -- that can all be used to obtain absorption spectra. In all these algorithms, we address the problem of targeting core-excited states of key relevance for XAS, saving significant computational resources in doing so. While previous works have proposed Green's function and QPE based algorithms for generic response functions \cite{bauer2016hybrid,endo2020calculation,kosugi2020construction,kosugi2020linear,cai2020quantum,sun2023quantum,kowalski2024capturing}, to our knowledge these have not been applied to X-ray absorption or to industrially-relevant problems for batteries; nor has it been explained how to target the core-excited states of interest in a quantum algorithm. After carrying out an asymptotic cost analysis of the algorithms and discussing their trade-offs in different scenarios, we conclude with an estimate of the resources needed to simulate the spectrum of an O-Mn molecular cluster in the Li-excess material Li$_2$MnO$_3$.
	
	\section{X-ray absorption spectroscopy\label{sec:xas-basics}}
	
	In XAS, highly energetic photons excite electrons from deep core orbitals around an absorbing atom into unoccupied valence orbitals (or the continuum, but we focus on the former). Strong localization of the core orbitals is what makes XAS highly sensitive to the immediate environment of an absorbing atom. Within the dipole approximation, for an incoming X-ray photon with frequency $\omega$, the absorption cross-section of XAS has the Kramers-Heisenberg form
	\begin{equation}
		\sigma_A(\omega) = \frac{4\pi}{3\hbar c}\omega \sum_{F \neq I} \sum_{\rho = x, y, z}\frac{\left| \bra{F}\hat{m}_{\rho}\ket{I} \right|^2 \eta}{((E_F - E_I) - \omega)^2 + \eta^2}.
		\label{eq:crosssection}
	\end{equation}
	Here $\ket{I}$ is the many-body initial state of the system with energy $E_I$, which we take to be the ground state since thermal effects are usually negligible at the X-ray energy scale. The dipole operator $\hat{m}_{\rho}$ excites core electrons into partially occupied electron shells, leading to non-zero overlap $\bra{F} \hat{m}_{\rho} \ket{I}$ only with excited states $\ket{F}$ of energy $E_F$ that possess a core hole --- the so-called \textit{core-excited} states. This is in contrast to lower-energy \textit{valence-excited} states, corresponding primarily to excitations of the electrons near the Fermi energy, as probed for instance with UV-visible spectroscopy (this is discussed further at the end of \cref{sec:xas-algos}). 
	
	The dipole matrix element $\bra{F} \hat{m}_{\rho} \ket{I}$ determines the strength of a particular absorption line, with $\eta$ the line broadening. Since the core orbitals in $\ket{I}$ are strongly localized, the dipole matrix element is only nonzero for similarly local contributions from $\ket{F}$. In this way, local electronic structure of the absorbing atom such as bonding, coordination, and oxidation has a strong effect on the X-ray absorption spectrum. Analyzing the spectrum then allows deducing this local electronic structure. And taken in reverse, since local structure overwhelmingly determines the spectrum, accurately simulating the spectrum only requires including an absorbing atom's immediate surroundings --- resulting in small problem sizes.
	
	\textit{Spectral fingerprinting.---}The dominant feature in an X-ray absorption spectrum is the \textit{absorption edge}, a sharp peak around the hydrogen-atom-like energy difference to the next principal electron shell. Atomic charge differences between different elements lead to appreciate separation of excitation energies. This means X-ray absorption edges for different atomic species are well-separated in energy, allowing to study their local electronic structure separately. There are several different edges within the XAS spectrum: the simplest edge to model is the $K$-edge, corresponding to $1s\rightarrow2p$ excitations. The next edge, corresponding to $2p\rightarrow3d$ excitations, is called the $L$-edge: to model it even qualitatively accurately, relativistic corrections such as spin-orbit coupling need to be taken into account. In this work we focus on the $K$-edge: however, our results can be straightforwardly extended to the $L$-edge spectra with the addition of scalar relativistic and spin-orbit corrections to the system Hamiltonian. 
	
	The simplest way to use XAS to determine the oxidation state of an atom in a material is through analyzing the portion of the spectrum nearest the absorption edge --- the X-ray absorption near-edge spectroscopy (XANES) part. XANES is uniquely determined by the oxidation state of the absorbing atom, a kind of spectral fingerprint. Matching the spectrum with known reference spectra from a database allows the determination of the oxidation state. 
	
	However, such an analysis can be limited in important ways. First, it is not always possible to find a reference spectrum --- the required phase could be unstable, or simply has not been measured and entered into a database. Second, reference spectra obtained from related materials could have other, non-oxidation-state differences that spoil the matching. Third, reference compounds are generally not tunable --- it is not possible to adjust lattice geometry or material composition at will. This last point is well-demonstrated in Ref. \cite{rana2014structural}: the complexity of matching the measured edges with the references in the case of Li$_2$MnO$_3$ neatly illustrates the challenge of only being able to use experimental reference spectra for fingerprinting analysis. 
	
	Because of all these limitations, ab initio simulations have become an integral part of XAS analysis. Given the local nature of XAS, only a small cluster immediately surrounding an absorbing atom needs to be modelled. An example of the clusters to be modelled in a Li$_2$MnO$_3$ material cathode at various stages of delithiation are pictured in \cref{fig:liexcess-clusters}.
	
	\textit{Classical methods.---}There is a broad variety of methods for modelling X-ray spectra: semiclassical approaches \cite{rehr1990scattering,rehr2000theoretical}, linear response time-dependent density functional theory (LR-TDDFT) \cite{stener2003time,ray2007description,besley2010time,liang2011energy,lopata2012linear,lestrange2015calibration,stetina2019modeling} and its variants like restricted open-shell configuration interaction singles (ROCIS) on DFT \cite{roemelt2013combined,kubas2018restricted}, Bethe-Salpeter equation approaches \cite{vinson2011bethe,li2019identifying,vinson2022advances}, and fully multireference, wavefunction-based methods \cite{roos1980complete,casanova2009restricted,maganas2014combined,yang2014multireference,pinjari2014restricted,guo2016simulations,sassi2017first,delcey2019efficient,maganas2019comparison,guo2020restricted,lee2023ab,casanova2022restricted}. Excited states in the XANES region, especially of materials containing transition metals, exhibit strong correlations: as such, truly accurate modelling of the spectra is at present achieved mainly with wavefunction-based methods. The state-of-the-art technique for ab initio XAS is the restricted active space (RAS) approach \cite{kasper2020ab}. In RAS, an extension of the usual complete active space (CAS) method, the active space of the chemical system is partitioned into three sectors: RAS1, where the maximum number of holes is fixed; RAS2, which allows any configurations; and RAS3, where the maximum number of electrons is fixed \cite{casanova2022restricted,casanova2009restricted}. The RAS scheme is naturally well-suited to XAS where core holes appear. 
	
	While RAS has been shown to be successful in a number of instances \cite{maganas2014combined,yang2014multireference,pinjari2014restricted,guo2016simulations,sassi2017first,delcey2019efficient}, it is still fundamentally limited by steep scaling with the number of orbitals. This can lead to appreciable deviations between the numerically simulated and experimentally observed spectra, for example in the case of the oxygen $K$-edge in magnetite, as described in Ref.~\cite{sassi2017first}. As each oxygen atom is surrounded in the first shell by four iron atoms, the minimal model for the local cluster should include at least 20 $d$-orbitals, which is already beyond the capabilities of RAS. Even in clusters where only a single transition metal ion is present, it is clear from the simulations presented in Ref.~\cite{sassi2017first} that there are deviations from experimental spectra that can likely be improved by access to larger active spaces. For instance, being able to include both the double-shell effect~\cite{andersson1992excitation,pierloot2003caspt2,sauri2011multiconfigurational,sharma2014low} and the bonding ligand orbitals~\cite{sharma2014low,maganas2019comparison} within the same calculation could lead to important accuracy gains, but in many cases this is beyond the present-day capabilities of RAS. In some contexts, the inclusion of a possibly large number of diffuse Rydberg orbitals could also be necessary~\cite{maganas2014combined}. 
	
	While DFT-based methods are more commonly used for XAS simulations, and modifications such as ROCIS/DFT \cite{roemelt2013combined,kubas2018restricted} have been developed and successfully used in a number of cases, their single-reference nature can leave them deficient in certain cases for calculations involving excited states of transition-metal-containing clusters. An example of this is shown in Ref.~\cite{kasper2020ab} for the FeCl$_6^-$ cluster, where DFT-based methods miss the multielectron excitations peaks that are picked up by more sophisticated wavefunction-based methods like RAS.
	
	\textit{Why quantum.---}There are several key reasons why XAS simulation is a more attractive application of early fault-tolerant quantum computers than ground-state energy estimation. 
	First, in practice, excited-state calculations have turned out to be more challenging for classical approaches like TDDFT and wavefunction-based methods than similar scale calculations targeting the ground state; meanwhile, quantum algorithms have similar costs for ground and excited states. 
	Second, the system sizes to be simulated, as determined by the ultralocal nature of XAS physics, are significantly smaller than are typically quoted for ground state calculations of materials --- a few-atom cluster is usually sufficient. This strongly reduces the qubit requirements of the algorithms. 
	Third, for such small clusters, ground-state preparation can typically be accomplished reliably with classical methods, removing one major roadblock for quantum algorithms. Excellent evidence that state preparation is achievable for the systems of interest in XAS simulation is available in the literature \cite{lee2023evaluating, fomichev2023initial}. The main conclusion of these studies is that for small clusters with only a few transition metal centers, in a reasonably-sized active space (up to 30-50 spatial orbitals), it is typically feasible to classically calculate and then implement on the quantum computer either the precise ground states, or states sufficiently close to them such that quantum algorithm efficiency is not seriously impacted by the remaining infidelity.
	Finally, accuracy demands for fingerprinting analysis in XAS are much less stringent than in ground-state energy calculations. Experimental resolution in XAS is rarely below 1 eV or at most 0.1 eV, which is appreciably higher than chemical accuracy (0.04 eV). 
	
	\section{Quantum Algorithm for X-ray absorption spectroscopy}
	\label{sec:xas-algos}
	In this section, we describe an end-to-end pipeline for performing XAS simulation on a quantum computer. We start by explaining the process of initial state preparation; we then describe three different quantum algorithms for the objective of calculating the cross section of interest, including a new time-domain approach based on Monte Carlo sampling. We also discuss how the aforementioned problem of core-excited states can be addressed, and finally give an estimate of the required resources for this task in the specific case of a Li-excess cathode material Li$_2$MnO$_3$.
	
	\textit{Initial state.---}
	The algorithm begins with the preparation of $\hat{m}_{\rho} \ket{I}$. We take the initial state $\ket{I}$ to be the ground state of a small, transition-metal containing cluster (some examples are discussed in \cref{sec:xas-apps}), and compute the best possible classical approximation using standard many-body techniques such as the density matrix renormalization group (DMRG) algorithm \cite{white1992density,white1993density,chan2011density,sharma2014low,lee2023evaluating,fomichev2023initial}. One of the appealing properties of the XAS problem is that often, for the clusters of interest, the ground state can be prepared extremely well using classical methods, but obtaining the XAS spectrum remains intractable. We then act on the approximate ground state $\ket{I}$ with the dipole operator $\hat{m}_{\rho}$ classically, and implement the result in the quantum register, using for example methods in Ref.~\cite{fomichev2023initial} or similar.
	Even if classical methods do not give precisely the ground state, they are likely to give a state with a high enough overlap with the ground state, so that QPE may be used to project on the ground state without much additional cost. In this case, we can implement the action of $\hat{m}_{\rho}$ using linear combination of unitaries (LCU) methods (see \cref{app:state-prep}). 
	
	The error in preparation of the initial state $\hat{m}_{\rho} \ket{I} / \| \hat{m}_{\rho} \ket{I} \|$ will enter the error of the final quantities of interest in an additive fashion; we assume it to be small and will not keep explicit track of it. We leave the rigorous consideration of its effect for future work: of special interest will be cases in which the prepared state has substantial overlap but the error is also not negligible.
	
	\textit{Green's function form.---}Given $\hat{m}_{\rho} \ket{I}$, the absorption cross-section in \cref{eq:crosssection} may be written in terms of the imaginary part of the following Green's function \cite{cai2020quantum}
	\begin{equation}
		\mathcal{G}_\rho(\omega) = \bra{I} \hat{m}_{\rho} \frac{1}{\hat{H} - E_I 
			- \omega + i\eta} \hat{m}_{\rho} \ket{I},
		\label{eq:g-freq-unnormalized}
	\end{equation}
	where $\hat{H}$ is the system Hamiltonian, $E_I$ is the energy of the initial state $\ket{I}$, $\omega$ the frequency of the X-ray photon, and $\eta$ the line broadening, which we will take to be roughly equal to experimental resolution ($\approx$ 1 eV). To recover the sum over excited states in the cross section, we simply insert a resolution of identity $\sum_F \ket{F} \bra{F}$ next to the resolvent fraction and use 
	\begin{equation}
		\text{Im}\left(\frac{1}{x+ i \eta} \right) = - \frac{\eta}{x^2+\eta^2}.
	\end{equation}
	This results in the following form for the cross section:
	\begin{equation}\label{eq:cross_section_green}
		\sigma_A(\omega) = - \frac{4\pi}{3\hbar c} \omega \sum_\rho \left[ \text{Im}\mathcal{G}_\rho(\omega) - \frac{ \| \hat{m}_\rho |I\rangle \|^2 }{-\omega + i\eta} \right],
	\end{equation}
	with the non-XAS $\ket{F} = \ket{I}$ term at the end that can be removed by subtracting from the dipole operator its expectation value $\hat{m}_{\rho} \rightarrow \hat{m}_{\rho} - \bra{I} \hat{m}_{\rho} \ket{I}$ in the initial state. We define a characteristic time $\tau = \mathcal{O}(\| H \|^{-1})$ and the \textit{normalized} Green's function 
	\begin{equation}
		G({\omega}) = \mathcal{G}(\omega) / \| \hat{m}_{\rho} \ket{I} \|^2,
	\end{equation}
	as the target quantity which needs to be evaluated to error $\epsilon$. We make this error dimensionless by forming the combination $\epsilon \tau^{-1}$.
	
\begin{figure}[t]
		\centering
		\includegraphics[width=0.7\linewidth]{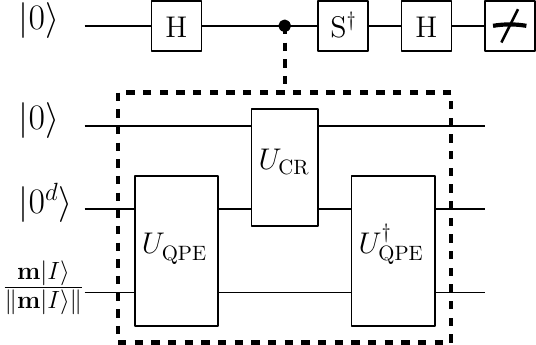}
		\caption{Quantum circuit implementing the frequency-domain Green's function approach to the simulation of X-ray absorption spectra. Having prepared the normalized initial state $\mathbf{m}\ket{I}/\|\mathbf{m}\ket{I}\|$, we use $d+1$ ancillas to perform QPE with a controlled rotation to invert the resolvent $(H - \omega + i\eta)^{-1}$. A global Hadamard test then yields the imaginary part of the Green's function $G(\omega)$ per choice of the gate $ S^{\dagger}$ on the Hadamard ancilla as shown in the circuit.}
		\label{fig:algo-circs-freqdomain}
	\end{figure}

	\textit{Frequency-domain.---}The first algorithm for obtaining the XAS spectrum directly calculates $G(\omega)$ in the frequency domain. This quantity can be readily obtained from a Hadamard test circuit \cite{cai2020quantum}, with different options for implementing the resolvent $(\hat{H} - \omega + i\eta)^{-1}$. The circuit (originally used in Ref. \cite{cai2020quantum}) is illustrated in \cref{fig:algo-circs-freqdomain}. It consists of three main unitaries: the first is performing an initial QPE that stores the energies of different components of the input state into a phase register. After that, the second unitary executes a controlled single qubit rotation based on the QPE outcomes, such that with outcome $0$ on the ancilla, it multiplies the register storing the eigenstate $\ket{E}$ by $(E - \omega + i\eta)^{-1}$ for each eigenvalue $E$. Finally, the third unitary uncomputes the phase register by doing a QPE with $-H$. Note that the implementation of all three of the operations, i.e.,~both QPEs and the controlled rotation, is controlled on the single Hadamard ancilla.
	The Hadamard test allows computation of the expectation value of $(\hat{H} - \omega + i\eta)^{-1}$ with a large enough number of samples. For details of implementation of these steps, see \cref{app:freqdomain-cost}.
	
	The asymptotic cost of this algorithm is as follows (see \cref{app:freqdomain-cost}):
	each run of the algorithm will require a number $\tilde{\mathcal{O}}((\eta\tau)^{-1} \log[1/(\epsilon\tau^{-1})])$ of queries to $e^{-iH\tau}$. Note that here we have taken the QPE accuracy to be $\mathcal{O}(\eta)$. The total number of repetitions for each value of $\omega$ will be 
	$\mathcal{O}((\epsilon \eta)^{-2})$. Lastly, the number of ancilla qubits required is $\mathcal{O}(\log(1/\eta) + \log (1/\epsilon_{\text{CR}}))) $, where $\epsilon_{\text{CR}}$ is the error in applying the conditioned rotation.
	
\begin{figure}[t]
		\centering
		\includegraphics[width=0.7\linewidth]{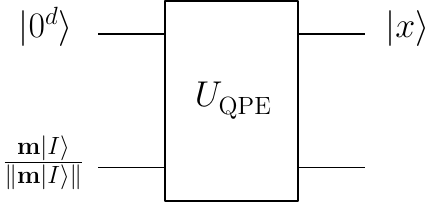}
		\caption{Quantum circuit implementing the QPE-based approach to X-ray absorption spectrum simulation. Performing QPE on the initial state $\mathbf{m}\ket{I}/\|\mathbf{m}\ket{I}\|$ directly reconstructs the absorption spectrum $\propto\text{Im}\,G(\omega)$.}
		\label{fig:algo-circs-qpe}
	\end{figure}

	\textit{QPE sampling.---}Another option to obtain the spectrum, having prepared the state $\hat{m}_{\rho}\ket{I}$, is to directly apply the standard QPE circuit (\cref{fig:algo-circs-qpe}), as also described previously in Refs.~\cite{kosugi2020linear, kowalski2024capturing}. QPE will return the excited state energy differences $E_F - E_I$ with each measurement: they will appear with probability given by $\left|\bra{F}\hat{m}_{\rho}\ket{I}\right|^2$. In this way, simply repeating rounds of QPE will build up the spectrum, starting with the strongest peaks. Roughly, we obtain the energies from the QPE outcomes and the overlaps $|\bra{F} \hat{m}_\rho \ket{I}|^2$ from the number of occurrences of each energy. 
	
	The result is then smoothed, using, for example, kernel density estimation \cite{wand1994kernel}. This essentially amounts to placing a kernel such as the Lorentzian kernel at the position of every outcome of QPE. We note that it is also possible to directly obtain a smoothed distribution from QPE with minimal overhead, as discussed in \cref{app:qpesampling-cost}.
	
	Asymptotic scaling analysis for this problem (see \cref{app:qpesampling-cost}) shows that the total number of repetitions of the circuit is $\mathcal{O}( (\epsilon\eta)^{-2} P(\omega)[1-P(\omega)])$, where $P(\omega)$ is the probability of obtaining $\omega$ as the QPE outcome. Each run requires $\tilde{\mathcal{O}}((\eta\tau)^{-1} \log[1/(\epsilon\tau^{-1})])$ queries to $e^{-iH\tau}$. The number of required ancilla qubits is $\mathcal{O}(\log[1/(\eta\tau)])$.
	
	\begin{figure}[t]
		\centering
		\includegraphics[width=0.7\linewidth]{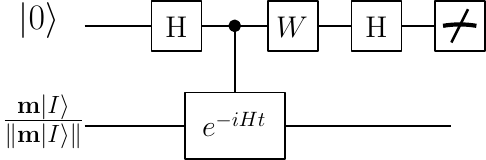}
		\caption{Quantum circuit implementing the time-domain Monte Carlo sampling based algorithm for X-ray absorption. Starting with the normalized initial state $\mathbf{m}\ket{I}/\|\mathbf{m}\ket{I}\|$, single-ancilla-controlled time evolution yields the real (imaginary) part of the Green's function in the time domain $G(t)$ via a Hadamard test for $W = I$ ($W = S^{\dagger}$). The evolution times are determined through Monte Carlo sampling: see the text for details.}
		\label{fig:algo-circs-timedomain}
	\end{figure}
	
	\textit{Time-domain.---}Finally, we present the third algorithmic approach to simulating X-ray absorption spectra. The scaled Green's function $G(\omega)$ may be equivalently written as a Fourier transform of the time-domain Green's function 
	\begin{equation}
		G(\omega) = \int dt  \; e^{i\omega t} e^{-\eta t} \tilde{G}(t).
	\end{equation}
	Similar to the frequency-domain approach, the Hadamard test circuit  in Fig.~\ref{fig:algo-circs-timedomain} gives direct access to $\tilde{G}(t)$. A discrete Fourier transform, computed on a classical computer, can then give the desired XAS spectrum. 
	
	We perform the Fourier transform by using a Monte Carlo sampling method, similar to the one introduced in Ref.~\cite{lin2022heisenberg}. To this end, we define the spectral measure
	\begin{equation}
		p(x) = \sum_{k=1}^K p_k \; \delta(x - \tau E_k),  \qquad x \in [-\pi , \pi),
	\end{equation}
	with $k$ running over all Hamiltonian eigenvalues $E_k$, $\tau = \mathcal{O}(\| H \|^{-1})$, $x = \tau \omega$, and $p_k$ denoting the probability of state $k$ in the normalized expansion of $\hat{m}_{\rho} \ket{I}$. In particular, we choose $\tau$ so that all eigenvalues of $H$ satisfy $-(\pi - \Delta) < \tau E_k < \pi - \Delta$, with $\Delta$ determined below. 
	The distribution $p(x)$ is defined above only in the range $[-\pi,\pi)$: we
	turn it periodic with period $2\pi$ for all real $x$ by copying it into all other intervals $[-\pi,\pi)+2n\pi$ for an integer $n$.
	We also define a periodic Lorentzian kernel (other kernels are also possible) as:
	\begin{equation}
		L_{\eta}(x) = \frac{1}{\pi} \sum_{n=-\infty}^\infty \frac{\eta \tau}{(x-2\pi n)^2 + (\eta\tau)^2},
	\end{equation}
	whose convolution with $p(x)$ results in
	\begin{equation}\label{eq:c_correaltion_def}
		C_\eta(x) = \int_{-\pi}^\pi dy \; p(y) \; L_\eta(x-y) = \frac{1}{\tau}\text{Im} \, G \left( \frac{x}{\tau} \right).
	\end{equation} 
	This, in effect, spreads every energy level as a Lorentzian with width $\eta$. If $\eta \ll \tau^{-1}$, the effect of periodic images is negligible. We also choose $\Delta$ to be an $\mathcal{O}(1/\log[1/(\epsilon\tau^{-1})])$ multiple of  $\eta\tau$. Note that $C_{\eta}$ is dimensionless and an error $\epsilon$ in $G$ is equivalent to an error $\epsilon\tau^{-1}$ for $C_{\eta}$.
	
	Similar to Ref. \cite{lin2022heisenberg}, the convolution in Eq.~\eqref{eq:c_correaltion_def} can be written in Fourier space as: 
	\begin{equation}
		C_\eta(x) = \frac{1}{2\pi} \sum_{j=-\infty}^\infty L_{\eta , j} \; \tilde{G} 
		(\tau j) \; e^{ijx},
	\end{equation}
	with $L_{\eta,j} = e^{-\eta \tau |j| }$ and we have noted that the Fourier transform of $p(x)$ for mode $j$ is $\tilde{G}(\tau j)$. 
	
	The sampling procedure is done as follows. First, a value of $J$ is drawn randomly according to a distribution given by weights proportional to $|L_{\eta , J}|$. Then two Hadamard tests are performed for time $\tau J$ in the circuit of Fig.~\ref{fig:algo-circs-timedomain}, resulting in random variables $X_J$ and $Y_J$ (they are taken equal to $+1/-1$ for $0/1$ outcomes of the Hadamard test); their expectation values recover $G(\tau J) = \mathbb{E}(X_J) +i  \mathbb{E}(Y_J).$ The variable $X$ requires $W = I$ and $Y$ requires $W = S^\dagger$.
	We then form the random variable
	\begin{equation}\label{eq:random_variable}
		C(x,J,X_J,Y_J) = L_\eta \ (X_J+ i Y_J) e^{i J x},
	\end{equation}
	where $L_\eta = \frac{1}{2\pi} \sum_{j=-j_{\text{max}}}^{j_{\text{max}}} L_j$ and   
	$j_{\max} = \mathcal{O}((\eta\tau)^{-1} \log[1/(\epsilon\tau^{-1})]) $.
	The expectation value of the above random variable is equal to $C_\eta(x)$ (up to error $\epsilon \tau^{-1}$) and thus proportional to the quantity of interest $G(\omega)$.
	
	The algorithm requires a number of runs asymptotically scaling as $\mathcal{O}( (\epsilon\eta)^{-2})$. 
	Furthermore, each run requiring on average $\mathcal{O}((\eta\tau)^{-1})$ calls to $e^{-iH\tau}$. The maximum runtime scales as $\mathcal{O}((\eta\tau)^{-1} \log[1/(\epsilon \tau^{-1})])$, however the \textit{average} runtime does not include the logarithmic factor. This gives an effective logarithmic improvement in single runs over the previous two algorithms.
	Only one ancilla qubit is required.

\begin{figure}[t]
		\centering
		\includegraphics[width=0.7\linewidth]{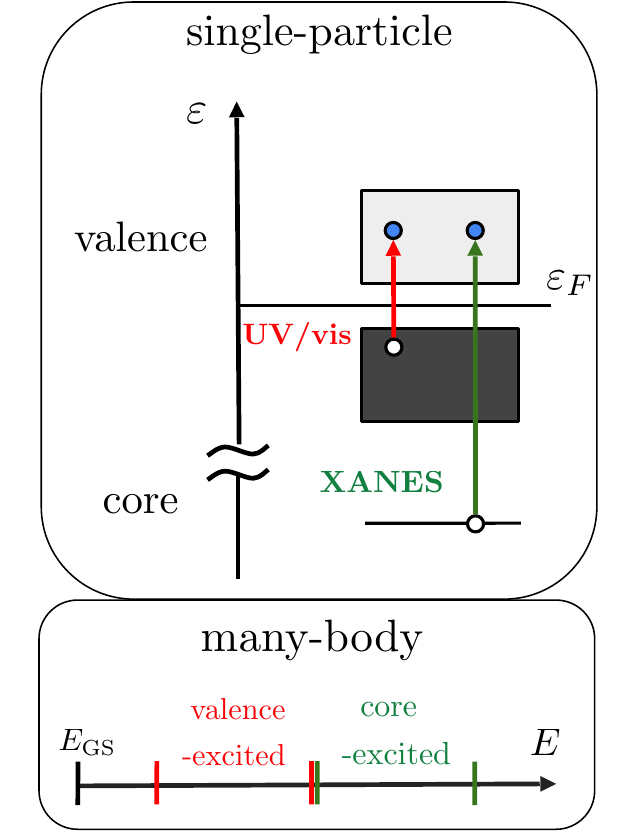}
		\caption{Illustration of the core-excited state problem. There are numerous low-lying excited states featuring mostly valence electron excitations: however, they appear on the same footing as the core-excited states which define the X-ray absorption spectrum.}
		\label{fig:core-excited-problem}
	\end{figure}

	\textit{Core-excited states.---}One final concern to address is the nature of the excited states probed by XAS. These are found high above the ground state by virtue of having a core hole. However, there are many (valence-excited) states in-between that are irrelevant to the X-ray absorption spectrum, but will still be computed by a naive application of some of the quantum algorithms above -- leading to significant cost overheads. This is illustrated pictorially in \cref{fig:core-excited-problem}.
	
	A low-cost method to avoid computing irrelevant excited states is to employ the core-valence separation approximation (CVS) \cite{cederbaum1980many,barth1981many,norman2018simulating,herbst2020quantifying}. It is based on the observation that the Hamiltonian matrix elements connecting core-excited determinants and valence-excited determinants are small: setting these matrix elements to zero would then exactly decouple the core-excited and valence-excited state manifolds.
	If at the same time we keep only terms in the dipole operator $\hat{m}_{\rho} = \sum_{ab} \hat{m}_{\rho,ab} c_{a}^{\dagger} c_b$ that are exciting the core electron, then initially the state $\hat{m}_{\rho} \ket{I}$ is placed into the core-excited subspace, and subsequently constrained to remain there due to lack of off-diagonal matrix elements. CVS is commonly employed in classical simulations, and has been shown to lead to errors on the order of 0.1 eV to at most 1 eV \cite{herbst2020quantifying} --- well-within experimental resolution of most XAS experiments. 
	Thus with CVS, \textit{any} quantum algorithm can avoid computing valence-excited states, at only a marginal loss of accuracy and no quantum costs. 
	
	A fully quantum alternative is to employ a filtering technique such as quantum eigenvalue transformation of unitary matrices with real polynomials (QETU) \cite{dong2022ground}. Applying filtering to the initial state $\hat{m}_{\rho}\ket{I}$ can localize its support in the expected X-ray spectral region (approximately determined through, for example, cheap mean-field methods), ensuring that non-core-excited states do not appear. Naturally, this method introduces non-trivial quantum costs, and carries with it a failure probability, but could be a more desirable option if the errors of CVS cannot be tolerated for a given application.
	
	\textit{Possible Scenarios.---} Each of the listed algorithms has some advantages and disadvantages. The time domain approach requires the least amount of resources in terms of ancillas and circuit depth, having a smaller average runtime per run by a logarithmic factor in $\epsilon$ compared to QPE sampling. Therefore, if one is restricted in terms of resources, as with an early fault-tolerant quantum computer, it might be the preferred method. 
	
	However, in terms of the number of repetitions, the QPE sampling method might be better, especially because it targets the more probable frequencies first. QPE sampling is also best when the situation corresponds to detecting a less probable spectral peak, i.e., when the height of the peak is small and the error is taken to be proportional to the peak height (for peak detection). 
	
	The frequency domain, on the other hand, has the highest overhead (performing QPE twice and the single qubit rotation), but allows one to directly target a specific frequency region. This can be beneficial when the errors introduced by the CVS approximation are too high but one still needs to avoid the computation of irrelevant, valence-excited states.
	
	\textit{End-to-end summary.---} Having described the steps of the algorithms in preceding sections, here we outline a step-by-step procedure for the most resource-friendly of them, namely the Monte Carlo time-domain algorithm with the CVS scheme.
	
	\begin{enumerate}
		\item Apply the CVS approximation to $H$ (this step is optional, only to reduce the cost):
		\begin{itemize}
			\item To account for the hole on a mean-field level, start from the Hartree-Fock orbitals obtained for a configuration with a hole in the core orbital \cite{cederbaum1980many}.        \item Having obtained the orbitals, set all two-electron integrals $\langle ij | kl \rangle$ that involve at least one core orbital to zero \cite{norman2018simulating}.
			\item Retain only terms in the dipole operator $\hat{m}_{\rho}$ that involve excitations from the core orbital.
		\end{itemize}
		\item Prepare the ground state $\ket{I}$:
		\begin{itemize}
			\item  If a classical method is capable of obtaining the ground state with an acceptable error, compute classically the action of the dipole operator $\hat{m}_{\rho}$, and implement the result $\hat{m}_\rho \ket{I} / \| \hat{m}_\rho \ket{I} \|$ on the quantum computer.
			\item If the best classical solution only has imperfect overlap with the ground state, implement that solution on the quantum computer and perform QPE to obtain the ground state, then use block-encoding for the dipole operator action (\cref{app:state-prep}).
		\end{itemize}
		\item Collect data from the Hadamard circuit of Fig.~\ref{fig:algo-circs-timedomain}:
		\begin{itemize}
			\item Classically sample the values of $J$, with a probability proportional to the Fourier transform of the kernel of interest, e.g.,~$L_{\eta,J}=e^{-\eta \tau |J|}$ for a Lorentzian kernel.
			\item Perform time evolution for a total time equal to $J\tau$ for each $J$, and collect the results of the Hadamard circuit measurement.
		\end{itemize}
		
		\item Calculate the cross section:
		\begin{itemize}
			\item Obtain the Green's function $G(\omega)$ in Eq.~\eqref{eq:c_correaltion_def} with the data from step 3 by calculating the mean of the random variable in Eq.~\eqref{eq:random_variable}.
			\item Use Eq.~\eqref{eq:cross_section_green} to calculate the cross section $\sigma_A(\omega)$. Here $\| \hat{m}_\rho |I\rangle \|$ is calculated either classically or through the statistics obtained from the block-encoding.
		\end{itemize}
	\end{enumerate}

	\section{Application to Li-excess cathodes}
	\label{sec:xas-apps}
	\begin{figure}[t]
		\centering
		\includegraphics[width=0.7\columnwidth]{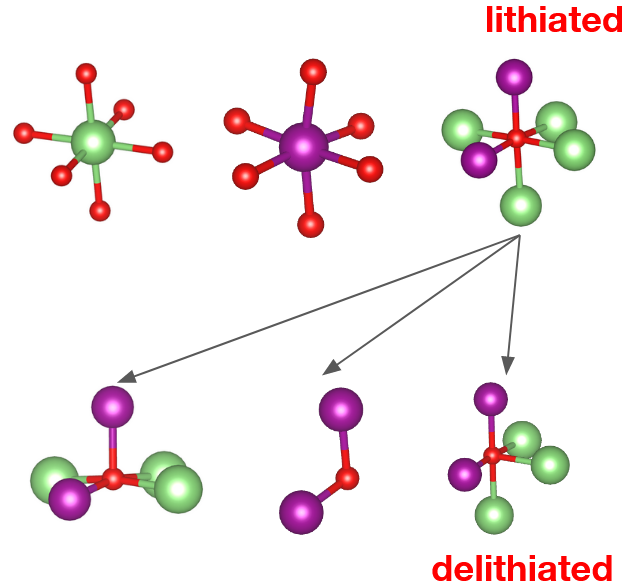}
		\caption{Examples of molecular clusters surrounding different absorbing atoms (green = Li, purple = Mn, red = O) that appear in Li$_2$MnO$_3$ cathodes in the lithiated and delithiated structures. The ultralocal nature of XAS means simulating the X-ray absorption of such clusters is sufficient for accurate modelling of the XANES portion of the spectrum.}
		\label{fig:liexcess-clusters}
	\end{figure}
	
	Having described quantum algorithms capable of producing an X-ray absorption spectrum, we now demonstrate the process on the example of the Li-excess material Li$_2$MnO$_3$. To probe the oxidation state of oxygen atoms, we focus on the $K$-edge absorption spectrum of oxygen. Upon the removal of lithium atoms during charging, a variety of oxygen-centered clusters could appear in the lattice structure: some of these are shown in \cref{fig:liexcess-clusters}. We use the geometry structure file from the Materials Project (Li2MnO3 (mp-1173926) from database version v2023.11.1 \cite{jain2013commentary})  as a starting point for delithiation and subsequent cluster extraction. For improved accuracy, starting configurations could be taken from ab initio molecular dynamics calculations \cite{li2019identifying}.
	
	We employ the cc-pVDZ basis set for all calculations. The minimal (valence) active space to model the spectrum for the largest O-centered cluster, Li$_4$Mn$_2$O, consists of the ten $3d$ orbitals of the Mn atoms, the four $2s$ valence orbitals of the Li and the three $2p$ O orbitals, together with the core $1s$ O orbital -- giving a total of 18 orbitals. There are several standard ways of generating this active space in the literature: for simplicity, we opt to use the automated approach called atomic valence active space (AVAS \cite{sayfutyarova2017automated}) as implemented in the electronic structure code PySCF \cite{sun2015libcint,sun2018pyscf,sun2020recent}. Given an overlap threshold, this method selects a subset of molecular orbitals that have the highest overlap with the chosen atomic orbital set (in this case, the Mn $3d$, Li $2s$ and O $1s$ and $2p$). For the particular geometry used, we adjusted the threshold until the application of AVAS yielded the expected minimal active space CAS(22e,18o). 
	
	This active space is already beyond the spaces typically employed in CAS and RAS calculations \cite{maganas2014combined,yang2014multireference,pinjari2014restricted,guo2016simulations,sassi2017first,delcey2019efficient}, motivating the use of a quantum algorithm. At the same time, calculations with only the minimum active space are known to be inaccurate, and in fact the prefereable active spaces could be even larger, further out of reach of classical approaches. One problem that leads to an increase in active space sizes is that for first row transition metals, the double-shell~\cite{andersson1992excitation,pierloot2003caspt2,sauri2011multiconfigurational,sharma2014low} correlation effect becomes important, meaning the $4d$ shells of Mn might also need to be included in the active space. Another addition could be other bonding ligand orbitals such as the Mn $3p$, as they could also affect accuracy. Other clusters such as the Mn-centered cluster MnO$_6$ also require large active spaces: a minimal active space for $K$-edge spectra for the cluster (b) in \cref{fig:liexcess-clusters} consists of 23 orbitals. The bottom line is that it would be highly desirable to be able to fit more orbitals into the active space, which is what the quantum approach promises to unlock.
	
	Having built the electronic Hamiltonian in the minimal active space CAS(22e,18o) for the top-rightmost cluster in Fig. \ref{fig:liexcess-clusters}, we can perform a rough estimation of resources required for the quantum algorithms. 
	Since resource estimation for QPE has been widely studied, we focus on the QPE sampling algorithm, using a qubitization-based approach for Hamiltonian simulation~ \cite{low2019hamiltonian}.
	Given that the cost of all algorithms is dominated by time evolution, we expect this estimate to be illustrative for all of them.
	We use the double factorization algorithm from Ref.~\cite{von2021quantum}, and employ PennyLane \cite{bergholm2018pennylane} to estimate the cost of QPE, which uses expressions from Ref.~\cite{lee2021even}. 
	This estimate, performed here for the accuracy of 1 eV (typical for XAS experiments), suggests about $900$ logical qubits and $10^8$ Toffoli gates for one sampling of the algorithm. Importantly, the large qubit overhead in this estimate is due to qubitization. If Suzuki-Trotter product formulas are used instead, this would reduce the qubit requirements by eliminating the need for ancillas. Indeed, under the Jordan-Wigner mapping, an active space of 18 spatial orbitals requires only 36 qubits. At the same time, focusing on the time-domain approach should allow for a reduction of circuit depth, as has been the case with similar algorithms previously~\cite{lin2022heisenberg}. Overall, while more work is needed to produce reliable and accurate estimates, preliminary studies indicate that obtaining XAS spectra for industrially-relevant systems could be within reach of early fault-tolerant quantum computers.
	
	\section{Conclusions}
	
	We argue that simulating the X-ray absorption spectrum of materials containing transition metals, for example Li-excess battery cathodes, is a promising application for quantum computing. 
	The ultralocal nature of the core-excited states requires simulating only the immediate shell of atoms surrounding the absorber site, significantly reducing the cost of quantum simulation compared to situations where entire supercells are needed. 
	At the same time, accessing the absorption spectrum is significantly harder for classical methods than calculations involving the ground state, making quantum approaches more competitive. 
	
	Out of the three algorithms proposed, the Monte Carlo sampling approach is the most hardware-friendly. 
	While the preliminary values we obtain for resource estimation are still high, they showcase how judiciously choosing the application can significantly reduce simulation costs. We anticipate that dedicated efforts to optimize quantum algorithms for this problem will be successful in further reducing cost of implementation.
	
	Given the limitations of classical methods for XAS, and the preeminent role that XAS plays in the analysis of oxidation states of battery cathodes, this work is the first step towards using future quantum computers to understand redox processes in Li-excess battery cathode materials and beyond. Knowledge of redox and other hard-to-measure chemical processes, as obtained from accurate ab initio simulations of XAS on a quantum computer, could then ultimately unlock a new series of advancements in materials development and improvement for high-capacity battery technology. 
	
	\bibliography{initial.bib}
	
	\appendix
	
	\section{Preparing the $\mathbf{m}\ket{I}$ state quantumly}
	\label{app:state-prep}
	
	While we expect that it is likely simpler and cheaper to implement the action of the dipole operator on the initial state classically and then directly prepare the state $\hat{m}_{\rho}\ket{I}$ in the quantum register, if that is not the case, we can still use LCU methods \cite{childs2012hamiltonian} for applying $\hat{m}_\rho$. While this will require postselection and discarding a number of samples, it can be done in parallel with all the qubits, and will not be a bottleneck. Even though the Hadamard test, which is a key component of two of the three algorithms, technically prevents the need for postselecting after applying $\hat{m}_{\rho}$, we still find it more beneficial to implement $\hat{m}_\rho$ and do postselection before going through the rest of the quantum algorithm. This is because doing postselection first prevents us from performing QPE in the cases that the dipole implementation failed; however if no postselection is performed, one essentially needs to perform QPE for a number of times equal to the cases where postselection succeeds and those in which it fails and this will be a decrease in cost for the postselected case.
	
	In the main text, we need to calculate and use the norm of the state $\hat{m}_\rho \ket{I}$, here we give details on how it can be calculated. If we have capabilities of computing the state classically this norm can also be calculated classically. However, if the state is implemented quantumly by first performing QPE and then using LCU techniques we can use the following procedure: first, we perform postselection to make sure that the desired state is implemented correctly. The probability of success is given by $\|\hat{m}_\rho \ket{I}\|^2 / \alpha_1^2$, where $\alpha_1$ is the LCU one-norm for implementation of $\hat{m}_\rho$. We can estimate this probability by the statistics of LCU success and we have $\alpha_1$ by construction. This leads to obtaining the value of $\|\hat{m}_\rho \ket{I}\|$.
	
	One last point is that since $\hat{m}_\rho$ is a single-fermion operator, we implement it in the basis in which it is diagonal and then write the LCU in terms of reflection operators that are made out of projectors onto eigenstates of $\hat{m}_\rho$.

	\section{Cost of frequency-domain algorithm}
	\label{app:freqdomain-cost}
	
	We calculate the cost of this algorithm in the following way:
	we implement the state $ [H-\omega+i\eta]^{-1} 
	\frac{\hat{m}_\rho | I \rangle}{\|\hat{m}_\rho |I\rangle \|}$ using the HHL algorithm \cite{harrow2009quantum}. First, decompose the state $\hat{m}_\rho | I \rangle  = \sum_k \beta_k |E_k\rangle$ in terms of eigenstates of the Hamiltonian $|E_k\rangle$. Performing QPE with the translated Hamiltonian $H-\omega$ and adding an auxiliary register, we obtain:
	\begin{equation}
		U_{\text{QPE}; \, \omega} |0\rangle \frac{\hat{m}_\rho | I \rangle}{\|\hat{m}_\rho |I\rangle \|} = \sum_k \beta_k |\lambda_k \rangle |E_k\rangle,
	\end{equation}
	where $|\lambda_k \rangle$ stores a binary representation of $E_k-\omega$.
	With the effect of the controlled rotation operator which we denote as $U_{\text{CR}}$, we obtain:
	\begin{equation}
		\sum_k \beta_k \left( \sqrt{1-\frac{c^2}{\lambda_k^2+\eta^2}} 
		|1\rangle + \frac{c}{\lambda_k+i\eta} |0\rangle\right)|\lambda_k \rangle |E_k\rangle,
	\end{equation}
	with $c$ a positive number that should be chosen such that $c \leq \sqrt{\lambda_k^2+\eta^2}$ for all $k$, in general the largest value of $c$ that could be used above is $\eta$ as there can be $\omega$ values close to an eigenvalue of $H$. We then uncompute the QPE register by performing $U_{\text{QPE}; \, \omega}^\dagger$:
	\begin{equation}
		\sum_k \beta_k \left( \sqrt{1-\frac{c^2}{\lambda_k^2+\eta^2}} 
		|1\rangle + \frac{c}{\lambda_k+i\eta} |0\rangle\right)|0 \rangle |E_k\rangle,
	\end{equation}
	Now if the unitary $U_{\text{QPE}; \, \omega}^\dagger U_{\text{CR}}U_{\text{QPE}; \, \omega}$ is used in a Hadamard test, we will be able to recover the following expectation value in terms of the probability $p_0$ of getting outcome 0 in the Hadamard register:
	\begin{equation}
		\begin{aligned}
			2p_0 - 1 &= \text{Im} \left(\sum_k \beta_k^2 \left\langle E_k \left| \frac{c}{\lambda_k+i\eta} \right|E_k \right\rangle \right)\\
			&= c \ \mathrm{Im}(G(\omega)).
		\end{aligned}    
	\end{equation}
	Note that this is because the QPE register and the controlled rotation qubit are initialized in states $0$. We have used the scaled definition of $G(\omega)$. 
	
	If we need to calculate the Green's function to error $\epsilon$, we need the probability with error $\epsilon c$. Using the Hadamard test error and the fact that we take $c=\eta$, the number of repetitions for each value of $\omega$ is then given by:
	\begin{equation}
		N_s^{\text{FD}} = \mathcal{O}\left( \frac{p_0(1-p_0)}{\epsilon^2 \eta^2}\right)
	\end{equation}
	
	\section{Cost of time-domain algorithm}
	
	\label{app:timedomain-cost}
	
	To estimate the cost of this algorithm, note that $L_\eta(x)$ is bound above by its maximum (denoted $L_{\eta}$) which obtains at $x = 0$: thus for the cost analysis, we can replace factors of $L_{\eta}$ with $\mathcal{O}(\frac{1}{\eta \tau})$. The variance of a single instance of $C(x)$ is given by:
	\begin{equation}
		\text{var} (C(x, J , X_J, Y_J)) \leq L_{\eta}^2  \, \mathbb{E}[|X_J|^2+|Y_J|^2] \leq 2 L_{\eta}^2.
	\end{equation}
	
	We define the following average:
	\begin{equation}\label{eq:C_average}
		\bar{C}(x) = \frac{1}{N_s}\sum_{k=1}^{N_s^{\text{TD}}} C(x, J_k, X_{J_k} , Y_{J_k}),
	\end{equation}
	with $N_s$ the number of samples.
	
	To obtain the Green's function $G$ to error $\epsilon$ (that has dimensions of Energy$^{-1}$), we need $C_{\eta}$ to error $\epsilon / \tau$. Thus we need to do the sampling in \cref{eq:C_average} $N_s^{\text{TD}}$ times that can be written in terms of $\epsilon$ as $\leq 2 L_{\eta}^2 \frac{1}{(\epsilon \tau^{-1})^2}$. This means that the total number of evolutions is given by:
	\begin{equation}
		N_s^{\text{TD}} = \mathcal{O}\left( \frac{1}{\eta^2} \frac{1}{\epsilon^2} \right)
	\end{equation}
	On the other hand, the expected number of times in a single run that one queries $e^{-iH\tau}$ is given by 
	\begin{equation}
		\begin{aligned}
			\mathbb{E}[|J|] &= \frac{1}{L_{\eta}} \sum_j |j| L_{\eta , j} = \mathcal{O} \left( \eta \tau \frac{1}{\eta^2\tau^2}  \right) \\
			&= \mathcal{O} \left( \frac{1}{\eta \tau} \right).
		\end{aligned}
	\end{equation}
	This means that the expected total number of times to query $e^{-iH\tau}$ scales as $\mathcal{O}( \frac{1}{\eta^3 \tau \epsilon^2 })$.

	Note that the maximum evolution time in a single run is given by $j_{\max}$, however the average evolution time in a single run is smaller by a logarithmic factor. This results in the average circuit depth to improve logarithmically  over the two other methods.

	We also note that this method can be used to calculate the energy distribution of any given state with Lorentzian (or other kernels for this matter) spreading of energy levels.
	
	\section{Cost of QPE sampling algorithm}
	\label{app:qpesampling-cost}
	
	To estimate the asymptotic cost of this algorithm, we first consider the case in which QPE is performed with the Lorentzian kernel. The state $\sum g(E) |E\rangle$ in the system register after performing QPE time evolutions takes the form
	\begin{equation}
		\frac{1}{2^{n/2}}\sum_{E,j} g(E) e^{-i \tau E j} |j\rangle |E\rangle.
	\end{equation}
	Here $n$ is the number of QPE qubits and we have $2^n = j_{\max}$.
	With the use of multi-qubit controls, we can arrive at the state:
	\begin{equation}
		\sum_{E,j} \ell_j \, g(E) e^{-i \tau E j} |j\rangle |E\rangle,
	\end{equation}
	with $\ell_j = L_{j}/ \sqrt{\sum L_j^2}$, with $L_j$ the Fourier transform of a kernel, for example for the kernel $\frac{1}{x+i\eta}$, we use $L_j = \theta(j) e^{-j \eta \tau}$.
	
	A QFT acting on the above state results in the following state:
	\begin{equation}
		\frac{1}{2^{n/2}} \sum_x \sum_{E,j} g(E) \ell_j e^{ij (x - E\tau)} |x\rangle |E\rangle,
	\end{equation}
	where $-\pi < x \leq \pi$ runs over discrete values. One recovers the integers $1,\ldots,2^n=j_{\max}$ by rescaling and translating all $x$ values as $\pi + 2^n x/(2\pi)$ (standard QPE integers).
	
	The probability of obtaining a value of $x$ now reads:
	\begin{equation}
		P(x) = \frac1{2^n} \sum_E |g(E)|^2 \left|\sum_j \ell_j e^{i j (x - E \tau)}\right|^2.
	\end{equation}
	We now specialize to the case of a Lorentzian kernel that was mentioned above, other kernels such as the original QPE kernel are similar. 
	We first note that with $j_{\max}$ chosen large enough, i.e.~an $O(1)$ (up to polylogarithmic) multiple of $\frac{1}{\eta \tau}$, the sum can be approximated as follows:
	\begin{equation}
		\left|\sum_j L_j e^{i j y}\right|^2 \approx \left|\frac{1}{y+i\eta}\right|^2 = \frac{1}{y^2 + \eta^2}.
	\end{equation}
	Note that we have $L_j$ instead of $\ell_j$ in the above equation. As a result of this, the value of the Green's function of interest, defined in Eq.~\eqref{eq:c_correaltion_def} can be evaluated as follows:
	\begin{equation}
		C_{\eta}(x) = P(x) (\eta\tau) \left(\sum L_j^2\right) 2^n.
	\end{equation}
	In order to obtain $C_{\eta}(x)$ with error $\epsilon \tau^{-1}$ ($\epsilon $ is dimensionful and the error in $G$), we need to perform measurements a number of times given by 
	\begin{equation}
		N_s^{\text{QPE}} = \mathcal{O}\left(\frac{ \text{var} \left[P(x) (\eta\tau) \left(\sum L_j^2\right) 2^n \right]}{\left( \epsilon \tau^{-1} \right)^2} \right).
	\end{equation}
	Noting that $\left(\sum L_j^2\right) = \mathcal{O} (\frac{1}{\eta \tau})$ and that $2^n = j_{\max} = \mathcal{O}(\frac{1}{\eta \tau})$, the total number of samples has a scaling:
	\begin{equation}
		N_s^{\text{QPE}} = \mathcal{O} \left( \frac{P(x) [1-P(x)]}{(\epsilon\tau^{-1})^2 } \ 2^{2n}\right),
	\end{equation}
	where we have used the variance of a Bernoulli distribution to find the variance of $P(x)$ which reads $P(x)[1-P(x)]$.
	
	A better approach to approximating the distribution will be to associate each frequency with a {\it range} of QPE outcomes rather than a single one, i.e.~taking the average probability in a window of length $\delta = \mathcal{O}(\eta \tau)$ around each rescaled frequency $\omega \tau$. One can now treat this as a new Bernoulli process and the probability of an outcome inside the above range reads:
	\begin{equation}
		\mathcal{P}(x , \delta) = \sum_{ x-\delta/2 < x' < x+\delta/2 } P(x').
	\end{equation}
	However, $C_{\eta}(x)$ can also be approximated as the average probability of getting any outcome within this range:
	\begin{equation}
		\begin{aligned}
			C_{\eta}(x) &= (\eta\tau) \left( \sum L_j^2 \right) 2^n \left[ \frac{\mathcal{P}(x,\delta)}{\delta/2^{-n} } \right]    \\
			&= (\eta\tau) \left( \sum L_j^2 \right) \frac{1}{\delta} \ \mathcal{P}(x,\delta),
		\end{aligned}
	\end{equation}
	where $\delta/2^{-n} $ is the number of points in the window of length $\delta$.

	In this way, we can use see that the variance of $\mathcal{P}(x,\delta)$ as $\mathcal{P}(x,\delta) [1-\mathcal{P}(x,\delta)]$. This results in the following result for the total number of samples:
	\begin{equation}
		\begin{aligned}
			N_s^{\text{QPE}} &= \mathcal{O}\left(\frac{ \mathcal{P}(x,\delta) [1-\mathcal{P}(x,\delta)]}{\left( \epsilon \tau^{-1} \delta \right)^2} \right)\\
			&= \mathcal{O}\left(\frac{ \mathcal{P}(x,\delta) [1-\mathcal{P}(x,\delta)]}{\left( \epsilon \eta \right)^2} \right),
		\end{aligned}
	\end{equation}
	where on the second row, we have used that $\delta = \mathcal{O}(\eta \tau)$. Note that the numerator is bounded by $1/4$. Upon approximating $\mathcal{P}(x,\delta)$ as $\mathcal{O}(C_{\eta}(x)\delta) = \mathcal{O}(C_{\eta}(x)\eta \tau) $ we arrive at:
	\begin{equation}
		\begin{aligned}
			N_s^{\text{QPE}} = \mathcal{O}\left(\frac{  C_{\eta}(x ) (\eta\tau) [1-C_{\eta}(x) (\eta\tau) ]}
			{\left( \epsilon \eta \right)^2} \right),
		\end{aligned}
	\end{equation}
	which expresses the required number of repetitions in terms of the distribution itself.
	
\end{document}